\documentclass{amsart}

\usepackage{amssymb}
\usepackage{amsmath}
\usepackage{multicol}
\usepackage{url} 
\usepackage[colorlinks]{hyperref}   
\usepackage{float}
\usepackage{kvdefinekeys}
\usepackage{graphicx}

\usepackage[margin=1in,a4paper]{geometry}

\usepackage{natbib}
\bibpunct{(}{)}{;}{a}{}{,}

\makeatletter
\def\@biblabel#1{\@ifnotempty{#1}{}}
\makeatother

\begin{document}


\title{GWSDAT - GroundWater Spatiotemporal Data Analysis Tool}
\author{Wayne R. Jones}
\author{Michael J. Spence}
\author{Adrian W. Bowman}
\author{Ludger Evers}
\author{Daniel A. Molinari}

\address[W. R. Jones, M. J. Spence]{Shell Global Solutions (UK), Shell Technology Centre Thornton, P.O. Box 1,Chester, CH1 3SH, United Kingdom}
\address[A. W. Bowman, L. Evers, D. A. Molinari]{School of Mathematics and Statistics, University Gardens, University of Glasgow, Glasgow, G12 8QQ}

\begin{abstract}
Periodic monitoring of groundwater quality at industrial and commercial sites generates large volumes of spatiotemporal concentration data. Data modelling is typically restricted to either the analysis of monotonic trends in individual wells, or independent fitting of spatial concentration distributions (e.g. Kriging) to separate monitoring time periods. Neither of these techniques satisfactorily elucidate the interaction between spatial and temporal components of the data. Potential negative consequences include an incomplete understanding of groundwater plume dynamics, which can lead to the selection of inappropriate remedial strategies. The GroundWater Spatiotemporal Data Analysis Tool (GWSDAT) is a user friendly, open source, decision support tool that has been developed to address these issues. Uniquely, GWSDAT applies a spatiotemporal model smoother for a more coherent and smooth interpretation of the interaction in spatial and time-series components of groundwater solute concentrations. GWSDAT has been designed to work with standard groundwater monitoring data sets, and has no special data requirements. Data entry is via a standardised Microsoft Excel input template whilst the underlying statistical modelling and graphical output are generated using the open source statistical program R. This paper describes in detail the various plotting options available and how the graphical user interface can be used for rapid, rigorous and interactive trend analysis with facilitated report generation. GWSDAT has been used extensively in the assessment of soil and groundwater conditions at Shell’s downstream assets and the discussion section describes the benefits of its applied use. These include rapid interpretation of complex data sets, early identification of new spills, detection of off-site plume migration and simplified preparation of groundwater monitoring reports - all of which facilitate expedited risk assessment and remediation. Finally, some consideration is given to possible future developments.

\end{abstract}


\maketitle



\section{Introduction}
\subsection{Background} \label{IntroBackGround}
Groundwater is water located beneath the earth's surface in soil pore spaces and in the fractures of rock formations. Environmental monitoring of groundwater is routinely conducted in areas where the risk of contamination is high and for protecting human health and the environment following an accidental release of hazardous constituents. Groundwater monitoring strategies are designed to establish the current status and assess trends in environmental parameters, and to enable an estimate of the risks to human health and the environment. It involves installing a network of monitoring wells to enable access to the water table across the site \citep{EPAPRacGuide}. Samples of groundwater are periodically collected from these wells and sent to an accredited laboratory for chemical analysis. The resulting spatiotemporal data set has to be reviewed, analysed statistically, interpreted, and the results presented to environmental regulators in a clear and understandable manner. 

The most basic method of level and trend evaluation involves investigating the time-series of groundwater constituent concentrations independently on a well by well basis. The more sophisticated spatial methods, typically, involve fitting a concentration trend surface (i.e. Kriging) to evaluate spatial pattern and trend \citep{GTS1,Gaus1}.  However, although spatiotemporal data lies at the heart of current research in statistical methods (see \cite{cressie-2011-book}), the lack of any readily available and `off the shelf' spatiotemporal modelling software tools has lead to the practice of independently applying spatial techniques to separate monitoring events  (e.g. \cite{RickerPlumeQuant}) or applying a single spatial model to a data set which has been consolidated over a time period (e.g. \cite{Maros1}). The joint modelling of both spatial and time elements in a single spatiotemporal modelling framework leads to a more coherent interpretation of site groundwater characteristics \citep{Eversetal2013}.

Whilst there are a range of freely available groundwater data analysis applications available the most sophisticated tend to be designed for large scale long term groundwater monitoring networks \citep{Maros1,Cameron1}. These have a relatively large initial data warehousing setup burden which may be viewed as a barrier to the more widespread use of advanced groundwater monitoring techniques to smaller more short term monitoring programmes. Similarly, whilst GIS applications (e.g. ArcGIS) have excellent visualisation tools for geographical interpretation they also have a high initial data setup cost, operator competence requirements, and perhaps surprisingly, only a limited number of geostatistical modelling techniques available.

\section{Software design and aims}

\subsection{Development aims}
To a large extent GWSDAT has been developed to address the barriers discussed in section \ref{IntroBackGround}. However, its most important aim is to provide a simple to use, but statistically powerful decision support tool to environmental engineers and practitioners who routinely report on the status of numerous groundwater monitoring sites. Such an application needs to be easy to setup yet powerful in its ability to objectively analyse and rapidly report on a groundwater monitoring site's characteristics. 

In common with many other environmental applications it was recognised that there would be a benefit in providing the software in an open and transparent manner because policy makers and environmental regulators generally prefer code and techniques which are fully transparent and supported by sound science \citep{openair1}. 


\subsection{Software architecture}\label{SoftwareGWSDAT}
GWSDAT has been designed to integrate with Microsoft Excel, a software routinely used by environmental engineers for storing and analysing environmental (e.g., soil and groundwater) data. The user entry point to GWSDAT is a custom built Excel Add-in menu (see top left of Fig. \ref{fig:docinfo}). 

The statistical engine used to perform geostatistical modelling and display graphical output is the open source statistical programming language R \citep{R}. 
The R project is used across a wide range of disciplines and has been adopted with eagerness by the environmental sciences community \citep{openair1}.
Members of the R community contribute statistical routines and functionality to this collaborative project by means of an open standardised package structure which can be downloaded and installed from \url{http://cran.r-project.org/web/packages/}. GWSDAT makes use of several of these packages which are all individually referenced in this article.  A Graphical User Interface (GUI) is provided via the R packages \emph{rpanel} \citep{rpanel1} and \emph{tkrplot} \citep{tkrplot1}, which obviates the need for training GWSDAT users in the R programming language.


\section{Software availability}
GWSDAT will be made freely available on the American Petroleum Institute website \url{http://www.api.org} - scheduled May/June 2013. 
It is designed and supported for all recent versions of Microsoft Office (Excel, Word and PowerPoint) running on Microsoft Windows. 
Installation is a simple two step procedure which, firstly, involves installing the R programming language, details of which can be found in \cite{R} and \url{http://www.r-project.org}. 
The second step is to download the GWSDAT installation files and install the Microsoft Excel add-in. No special hardware is required to run GWSDAT other than a standard desktop computer running on a windows operating system. Two example data files are provided for training and demonstration purposes.

\section{Data input}
\subsection{Background} \label{GWBackground}
Before describing the application of GWSDAT in more detail it is necessary to give a brief explanation of the nature of groundwater monitoring data. In general, routine sampling of a monitoring well involves measuring the water level elevation and taking a sample of the groundwater which is subsequently sent for laboratory analysis to ascertain the dissolved concentration of a prescribed set of solutes (e.g. Toluene, Benzene).  If the concentration is deemed lower than that which could be detected using the method employed by the laboratory then it is classified as a `non-detect'. In such circumstances the laboratory quotes the detection threshold concentration value below which the solute could not be detected.

An additional important consideration for petroleum hydrocarbon applications is the presence of a layer of Non-Aqueous Phase Liquid (NAPL), such as gasoline or diesel, on the surface of the water table. This circumstance often arises when the amount of contamination is sufficient to exceed the natural solute level of groundwater. Samples containing NAPL are not often sent for a full chemical analysis (unless performing NAPL forensics) because the levels of solute concentrations are too high for the traditional laboratory methods which are geared towards lower concentrations. Hence, NAPL data poses the challenge of how to handle unspecified high solute concentration values and identify trends in NAPL layer thickness.

\subsection{Input data format}
\begin{figure*}[h!]
\begin{center}
\includegraphics[width=400pt,angle=0,clip=0]{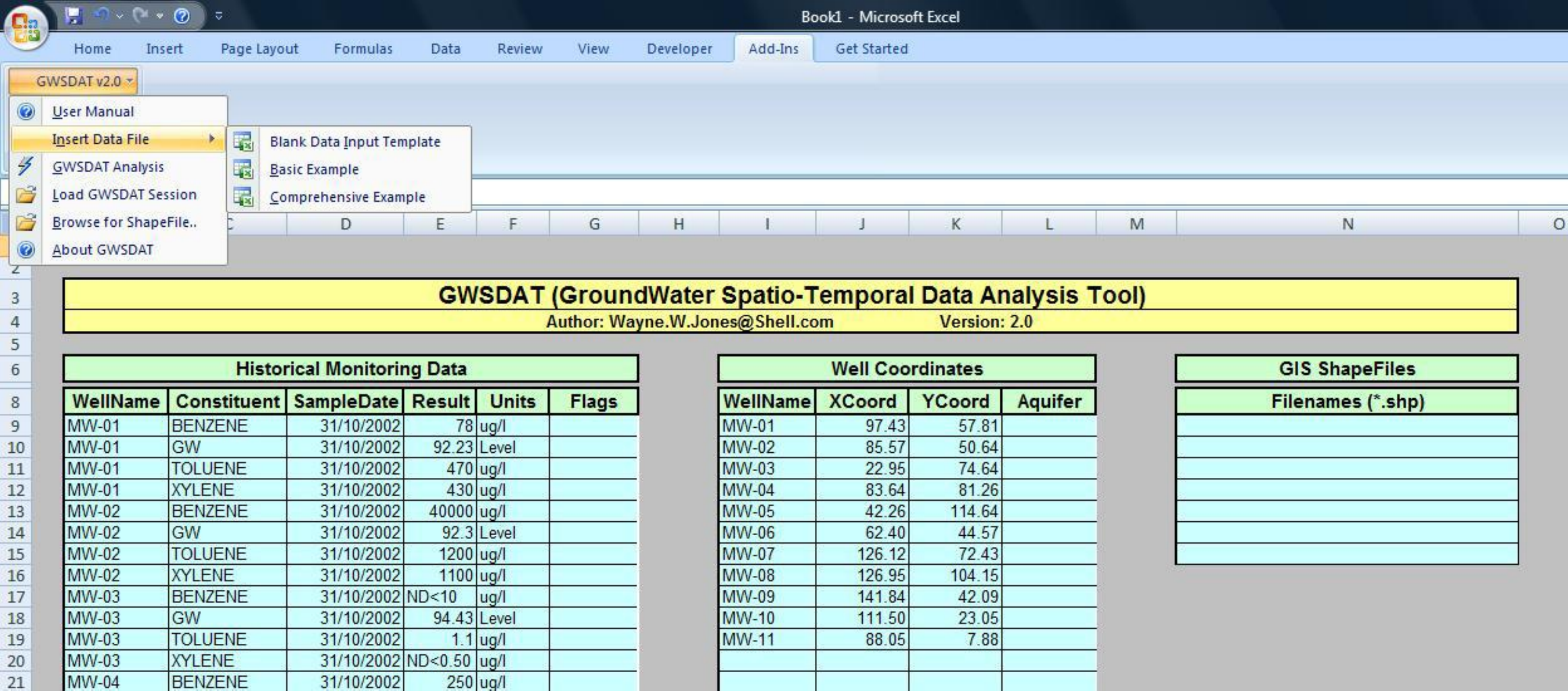}
\caption{GWSDAT example data input template. The \emph{Historical Monitoring Data} table captures the concentration data, groundwater levels and, if present, NAPL thickness. 
The \emph{Well Coordinates Table} stores the location of the monitoring well. The GWSDAT add-in menu is displayed at the top left. }
\parbox{460pt}{}
\label{fig:docinfo} 
\end{center}
\end{figure*}

Groundwater monitoring data is entered into GWSDAT by means of a simple standardised Microsoft Excel input sheet (Fig. \ref{fig:docinfo}). There is no requirement to gather any data that would not have already been recorded in a standard groundwater monitoring data set. The following summarises the GWSDAT data input format but the reader is referred to the user manual for a full and detailed explanation of GWSDAT data input specification.

Each row of the \emph{Historical Monitoring Data} table (left hand table in Fig. \ref{fig:docinfo}) corresponds to a unique combination of well id, sampling date, aquifer zone, solute name and concentration.  Non-detect solute data is entered using the notation `ND$<$X', where X represents the laboratory reported detection threshold concentration. If present, NAPL thickness data is also entered in this table using the constituent name `NAPL' with an appropriate unit, e.g. metres, mm. Optionally, groundwater level data is entered here (using the constituent name `GW') as an elevation above a common datum, e.g. metres or feet above sea level or some other common reference height.

The \emph{Well Coordinates} table (middle table in Fig. \ref{fig:docinfo}) stores the coordinates of the groundwater monitoring wells. 
Any arbitrary coordinate system with an aspect ratio of 1 can be used,  i.e. a unit in the x-coordinate is the same distance as a unit in the y-coordinate.

The third optional \emph{GIS Shapefiles} table can be populated with file locations of GIS shapefiles \citep{ESRIShapefile1998} for use as basemaps or site plans. 
Two GWSDAT input data sets of varying complexity (basic and comprehensive) are included with the software to serve as both an example of the GWSDAT data input format and provide a quick way of getting started. 

\subsection{Data processing}
On initiation of a GWSDAT analysis, the user is asked to select from a variety of data processing options including the handling of non-detects and, if present, NAPL. In accordance with the common convention, the default option is to substitute the non-detect solute concentration data with half its detection limit. For a more conservative choice, the user can select the alternative of substitution with the full detection limit. If NAPL is present the user is prompted to substitute NAPL data points with site dataset maximum observed solute concentrations. This option is to provide a more realistic picture of the area of impacted groundwater (high concentrations) in the event that NAPL in wells prevents direct measurement of solute concentrations as discussed in section \ref{GWBackground}. 
The data processing step is concluded with a series of data validation procedures to check for common data input errors.

\section{Graphical user interface}

\subsection{Introduction}
\begin{figure*}[ht!]
\begin{center}
\includegraphics[width=320pt,angle=0,clip=0]{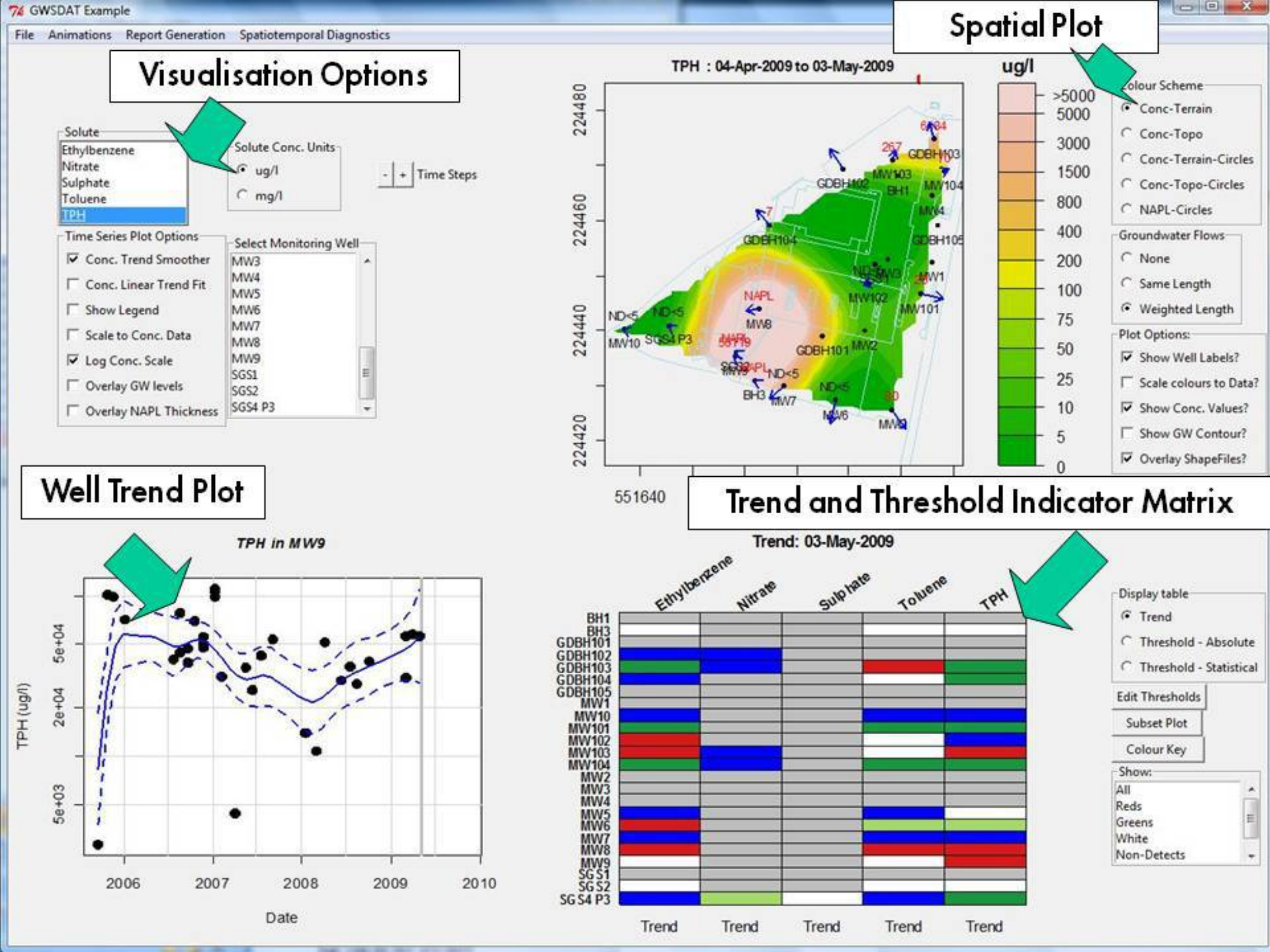}
\caption{The GWSDAT graphical user interface is a stand-alone, point and click, Graphical User Interface (GUI) which enables the user to perform a rapid, rigorous and interactive analysis of trends in the time-series, spatial and spatiotemporal components of the data.}
\label{fig:GWSDAT GUI} 
\end{center}
\end{figure*}

In the interests of user-friendliness and productivity the results of a GWSDAT analysis are interrogated and interpreted through the GWSDAT user interface (see Fig. \ref{fig:GWSDAT GUI}).
It includes a wide range of different plots for the visual inspection of groundwater monitoring data. The objective assessment of trend is achieved by the application of statistical smoothing models described in  \ref{ModellingDescription}. The following sections describe the individual components of the GWSDAT user interface in more detail.

\subsection{Well trend plot}\label{SectWellTrendPlot}

\begin{figure}[h!]
\begin{center}
\includegraphics[width=250pt,angle=0,clip=0]{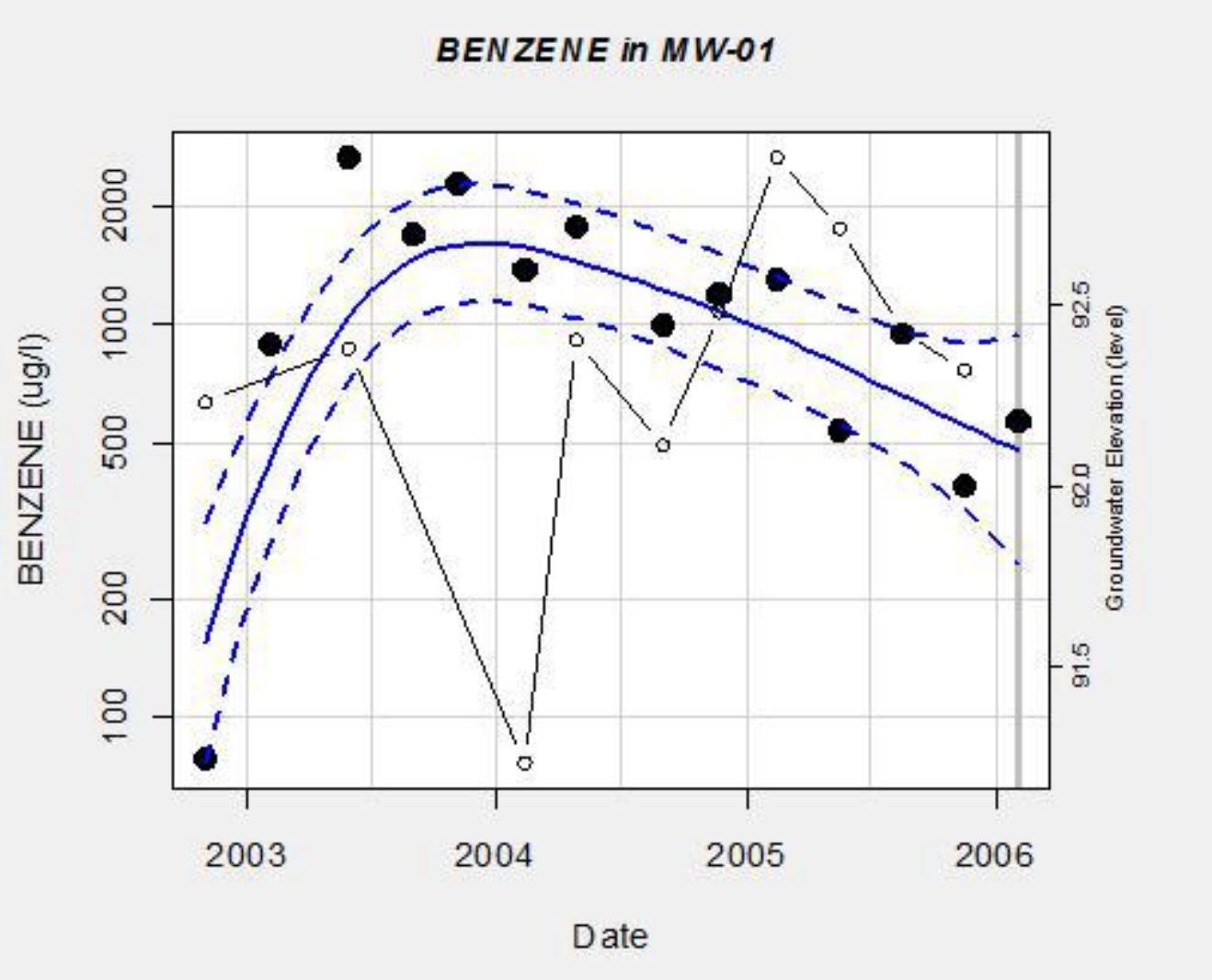}
\caption{Example of the GWSDAT well trend plot. The black solid circles represent observed concentration values. Overlaid in the solid blue line is a local linear regression model fit with 95\% confidence interval, shown as dashed blue lines. The open circles joined by solid line represent groundwater elevation measurements which are read off from the right hand axis. }
\label{fig:welltrendplot} 
\end{center}
\end{figure}

The well trend plot (see Fig. \ref{fig:welltrendplot}) enables the user to investigate time series trends of solute concentrations and groundwater level in individual wells. 
Sampled concentration values are displayed using orange circles for non-detect data and black solid circles for detectable data. 
The user can choose to overlay a linear (or log-linear) regression model fit and use the non-parametric Mann-Kendall approach to trend detection via the R package \emph{Kendall}  \citep{Kendall1}.
Although this approach is widely used in environmental sciences (e.g. \cite{HirschEtal1982,HelselHirsch2002}) its major weakness is that it can only detect monotonic trend and in response GWSDAT adopts an additional methodology. 
The solid blue line in Fig. \ref{fig:welltrendplot} displays the estimate (together with a 95\% confidence interval) of the mean trend level according to a local linear regression model fit described in \ref{smregressCalc}.
This non-parametric model smoothing technique is not constrained to be monotonic and can change direction as is clearly illustrated in the figure. The trend between two points in time is, informally speaking, deemed statistically significant if the associated confidence intervals do not overlap.

For evaluating the impact of changing (perhaps seasonal) water table conditions groundwater elevation data can, optionally, be overlaid in this plot. The time series of observed groundwater level is represented by open circles joined by a black solid line see and the values read off from the right hand axis (see Fig. \ref{fig:welltrendplot}). If present, NAPL thickness data can also be displayed in a similar manner.

\subsection{Trend and threshold indicator matrix}\label{SecTrendthresholdindmat}

\begin{figure}[h!]
\begin{center}
\includegraphics[width=300pt,angle=0,clip=0]{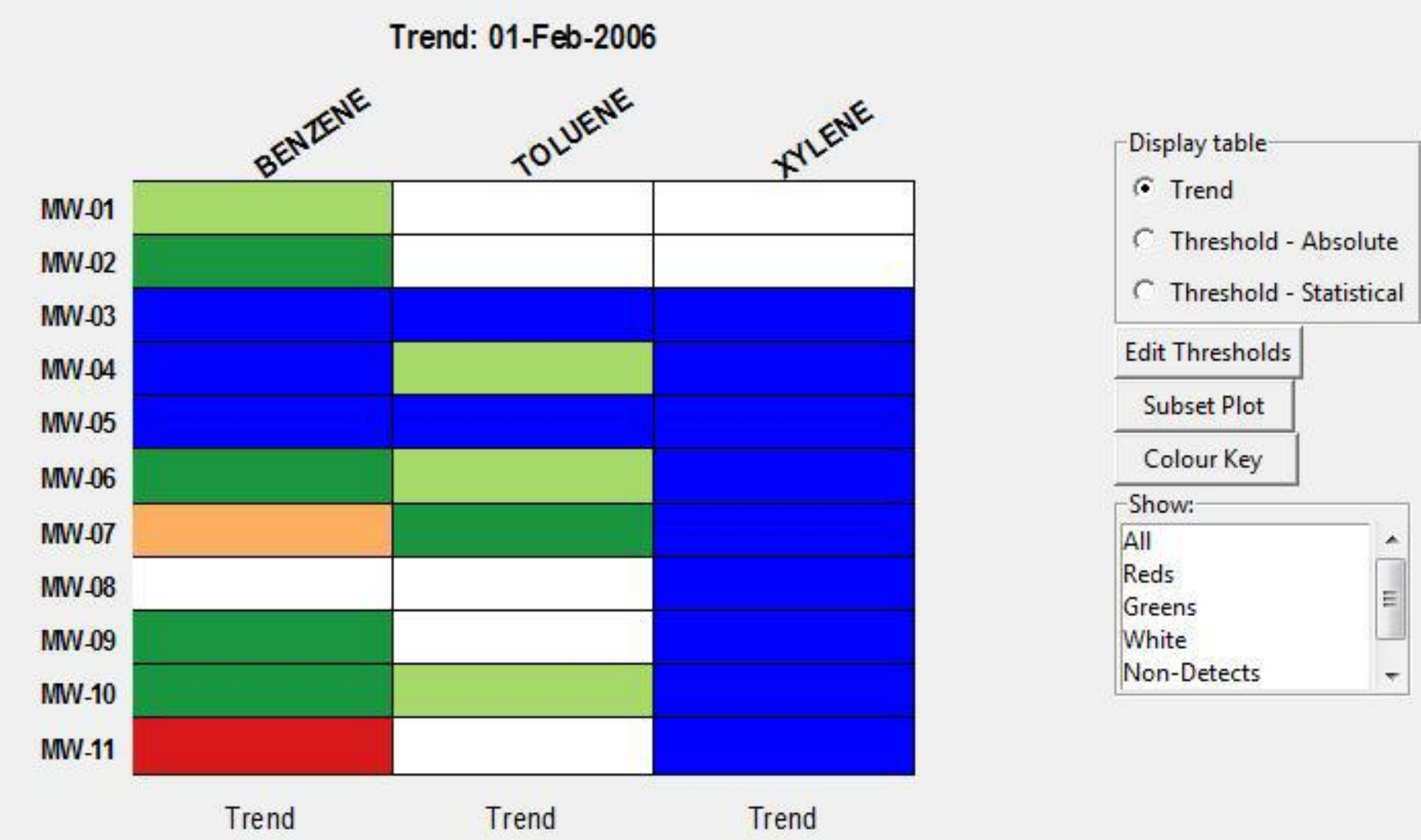}
\caption{Example of the GWSDAT trend and threshold indicator matrix. The rows represent monitoring wells, and columns represent the different solutes. Each cell is colour coded to represent increasing (reds), stable (white) or decreasing (greens) trends in solute concentrations. Blue cells represent non-detect data and if there is insufficient data the cell is coloured grey.}
\label{fig:TrendThresholdIndMatplot} 
\end{center}
\end{figure}

The trend and threshold indicator matrix is a summary of the level and time-series trend in solute concentrations at a particular time-slice of the monitoring period. 
The rows correspond to each monitoring well and the columns correspond to the different solutes. The date of the time-slice is displayed at the top of the plot and also indicated by a vertical grey line in the well trend plot (see Fig. \ref{fig:welltrendplot}). The user can select between the options of displaying `Trend', `Threshold - Absolute' or `Threshold – Statistical'.

When `Trend' is selected the cells are coloured to indicate the strength and direction of the current trend as assessed by the instantaneous gradient of the well trend smoother (see section \ref{SectWellTrendPlot}) at the current time-slice . 
White cells indicate a generally flat trend whilst reds and greens indicate strong upward and downward trends, respectively. In the event that the trend cannot be calculated (e.g. no data) then the corresponding cell is coloured grey. Blue
cells represent non-detect data.

When `Threshold – Absolute' is selected the cells are coloured according to whether the observed current solute concentrations are below a user specified threshold value, such as a risk-based remedial objective. The cells are coloured red if the current solute concentration is above the threshold value and green otherwise. `Threshold – Statistical' is similar but only colours the cell green if the
upper 95\% confidence interval of the well trend smoother (see section \ref{SectWellTrendPlot}) is below the threshold value.

\subsection{Spatial Plot}\label{SecSpatPlot}

\begin{figure}[h!]
\begin{center}
\includegraphics[width=300pt,angle=0,clip=0]{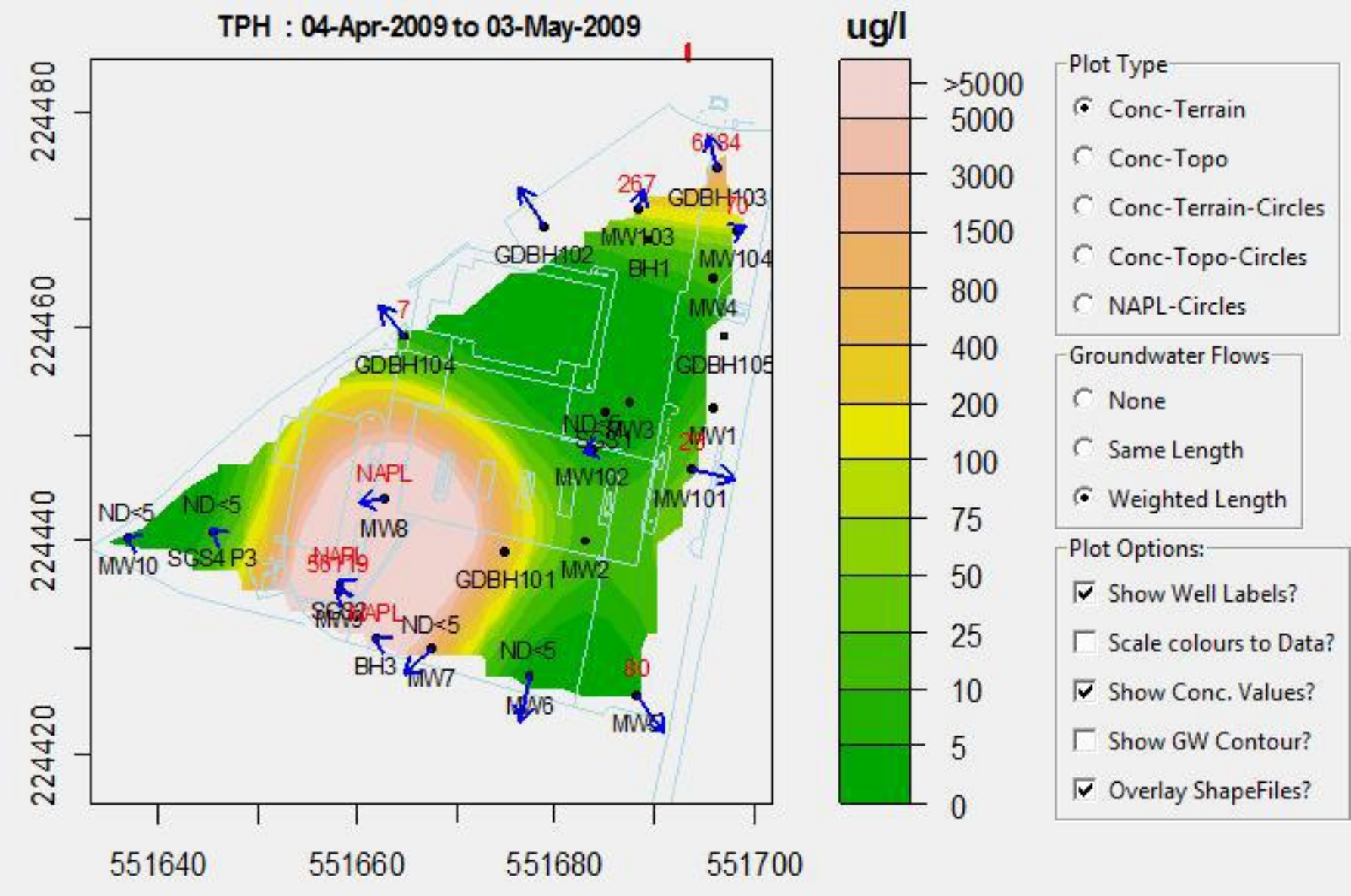}
\caption{Example of the GWSDAT spatial plot. The location of the named monitoring wells are depicted with black solid circles. Detect data or NAPL is displayed in a red font and non-detect in a black font above the wells. Blue arrows indicate vectors of estimated groundwater flow velocity. Spatiotemporal solute concentration smoother predictions are superposed using the colour key on the right. GIS shapefile data is overlaid using light blue lines.}
\label{fig:Spatialplot} 
\end{center}
\end{figure}

The GWSDAT spatial plot (see Fig. \ref{fig:Spatialplot}) is for the analysis of spatial trends in solute concentrations, groundwater flow and, if present, NAPL thickness.
It displays the locations of the named monitoring wells together with sample solute concentration values collected within the date interval displayed at the top of the graphic.
If desired, the major site features  (e.g. roads, fuel tanks), supplied in a GIS shapefile format, can be overlaid on the spatial plot as light blue lines.
As the user increments forwards and backwards through the monitoring history, using the `+' and `-' \emph{Time Steps} buttons, the spatial plot is updated. 

The estimated groundwater flow direction and magnitude is depicted with blue arrows calculated using the method described in \ref{GWCalc}. Additionally, it is possible to overlay a contour plot of groundwater elevation. This is achieved by drawing isopleths through a fitted local polynomial regression model fit implemented using the R function \emph{loess} - a 2D variant of the local linear regression method explained in \ref{smregressCalc}.

The spatial distribution of solute concentration is estimated by taking a time-slice through the spatiotemporal concentration smoother (discussed further in section \ref{SpatioTempSect}).
The model predictions are superposed on the spatial plot with a user-specified colour key located to the right of the plot. 
Alternatively, if no model based predictions are required, the concentration smoother can be replaced by size scaled colour coded circles representing the magnitude of sampled solute concentration values.
If NAPL is present, the additional `NAPL-Circles' option is available which displays NAPL thickness measurements at the monitoring well locations using a similar circle based representation, i.e. a bubble-plot.

The spatial plot uses the R packages, \emph{sp}  \citep{sp}, \emph{splancs} \citep{splancs} and \emph{maptools} \citep{maptools}.

\subsection{Spatiotemporal trend analysis} \label{SpatioTempSect}
One of GWSDAT's most unique features is that the spatial and temporal components of the solute concentration data are modelled jointly in a single modelling framework described in \ref{STMathSmoother}. 
The simultaneous statistical smoothing of both spatial and temporal components provides a clearer and more insightful interpretation of the groundwater monitoring site solute characteristics than would otherwise be gleaned from analysing these two components separately. However, it is not an inconsiderable challenge to effectively communicate the 3-dimensional nature of spatiotemporal trend through a 2-dimensional medium of a computer monitor. Furthermore, there is an additional constraint that the output from a GWSDAT analysis is commonly used in paper-based non-interactive reports submitted to environmental regulators. 
For this reason, GWSDAT communicates spatiotemporal trend through automatic plotting of the full temporal sequence of spatial plots (see section \ref{SecSpatPlot}). This animation based approach provides a `movie' clearly illustrating how both the spatial and temporal distribution of historical groundwater solute concentrations have changed over the monitoring period. 

The `animations' menu located at the top-left of the GWSDAT user interface (Fig. \ref{fig:GWSDAT GUI}) provides three different methods for generating animations. The first method plots and records the full sequence of spatial plots in an R graphics window. The user can toggle forwards and backwards through the sequence of spatial plots using the `Page Up' and `Page Down' keyboard buttons. 
The second method is identical but additionally generates a Microsoft PowerPoint slide-pack of the full sequence of spatial plots. The third method uses the R package 
\emph{animation}  \citep{animationRpackage}, to generate a html animation page (with controls) of spatial plots in the user's internet browser. The html animation can be viewed independently of GWSDAT, and hence provides an excellent dynamic media for communicating results to individuals who do not have direct access to GWSDAT.

\subsection{Report Generation}

\begin{figure}[h!]
\begin{center}
\includegraphics[width=320pt,angle=0,clip=0]{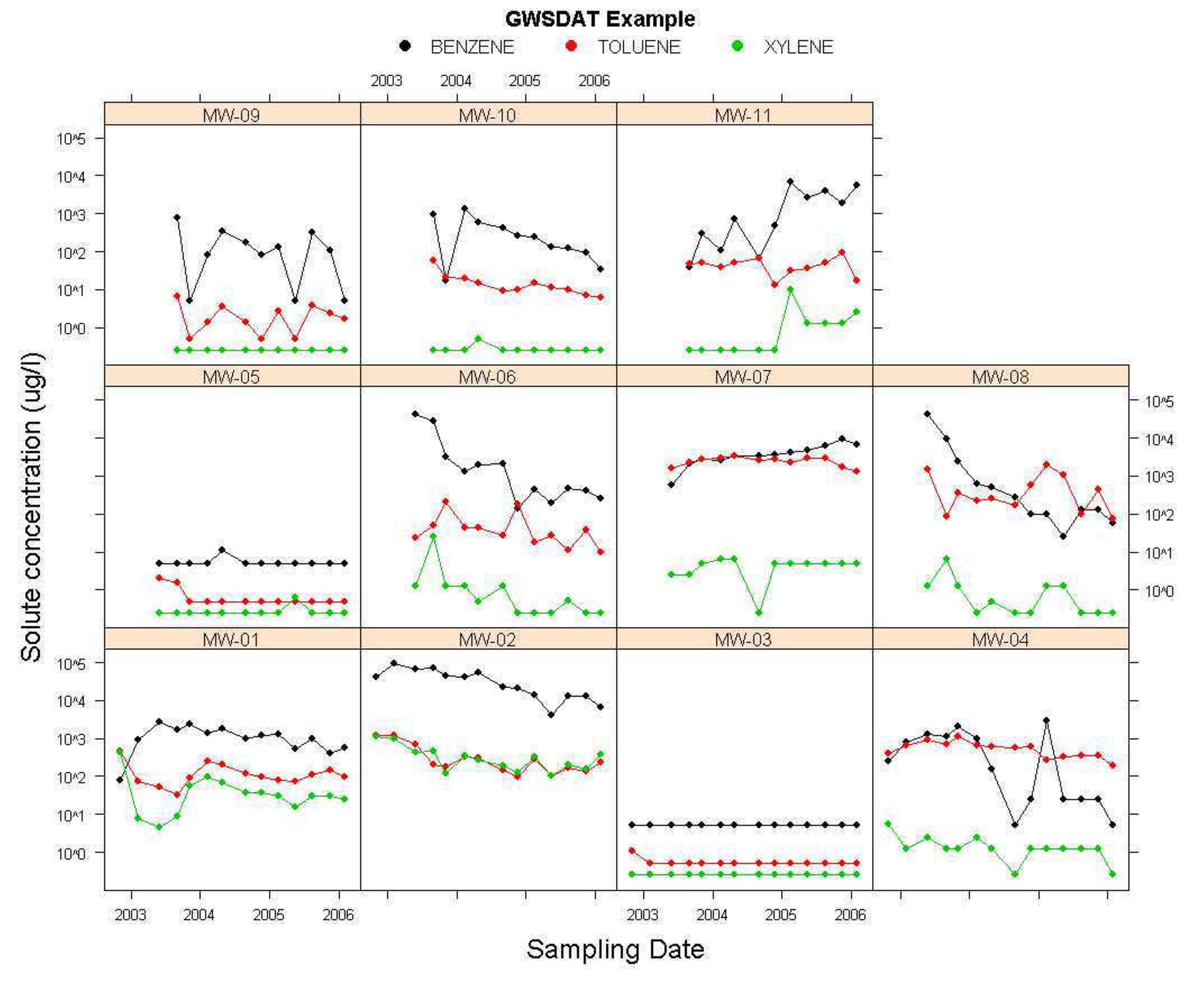}
\caption{Example of the GWSDAT Well report plot. The colour key at the top identifes each solute and the name of each well is displayed in a banner at the top of each of the individual time series graphs. This clearly illustrates the correlation in time series trends across the different solutes. }
\label{fig:WellReportingplot} 
\end{center}
\end{figure}

By left-clicking on any of the GWSDAT user interface plots, an identical but expanded plot is generated in a separate R graphics window. Plots can be saved to a variety of different formats including `jpeg', `postscript', `pdf', `metafile'.   Alternatively, with a single click of a mouse, plots and sequences of plots (e.g. spatiotemporal animation described in section \ref{SpatioTempSect} can be diverted directly in to Microsoft Word or PowerPoint. This functionality, implemented using the R package \emph{RDCOMClient} \citep{RDCOMClient}, enables the user to interactively compile a site groundwater monitoring report in an expeditious manner. 

Additional report generation functionality include the `Well Reporting' procedure, implemented using the R package \emph{lattice} \citep{lattice}, which generates a matrix of graphs displaying time series solute concentration values on a well by well basis (see Fig. \ref{fig:WellReportingplot}). 
This plot can be used to very concisely summarise the time series trends in the complete set of solutes and monitoring wells. A similar report procedure `GW Well Reporting' also allows for the overlay of the time series in groundwater elevation at each well. Finally, the `Latest Snapshot' procedure generates a sequence of plots (to PowerPoint if required) which reports on the most recent trends. This includes the latest spatial plot for each solute together with the most recent three variants of the `Trend and Threshold Indicator Matrix' plot described in section \ref{SecTrendthresholdindmat}.

\section{Discussion}
Environmental risk-based management decisions are often based on limited understanding of groundwater data, and relatively limited statistical analysis of that data.  GWSDAT has been designed and developed as a user-friendly, interactive, trend analysis tool for distilling the maximal information from such groundwater monitoring data sets. The application has been used operationally in the monitoring and assessment of Shell's global downstream assets (e.g. refineries, terminals, fuel stations) for a period of over 4 years. Graphical output generated from GWSDAT is routinely included in reports submitted to environmental regulators. Environmental engineers using GWSDAT have reported numerous benefits: 

\begin{itemize}

\item Rapid interpretation of complex data sets for both small and large groundwater monitoring networks.
\item Earlier identification of new spills or off-site migration.
\item Reduced reliance on engineered remediation through increased use of monitored natural attenuation remedies, where groundwater data analysis supports its effectiveness. 
\item Earlier closeout of sites in needless long-term monitoring and/or remediation.
\item Simplified preparation of groundwater monitoring reports.

\end{itemize}

\section{Future Developments}
The major area for future development is the addition of new capabilities to GWSDAT.  The assessment of solute plume stability is currently carried out by visually inspecting the evolution of the spatiotemporal solute concentration smoother. Feedback from users has highlighted the need for additional quantitative  tools to supplement this graphical method. Development is currently underway to incorporate plume mass balance tools,  such as those proposed in \cite{RickerPlumeQuant}, to automatically estimate plume characteristics such as area, total mass and centre of mass. The inspection of these quantities over the monitoring period will more objectively illustrate whether the plume is moving and if the plume is growing, shrinking or stable. 

Future versions of GWSDAT will use spatiotemporal model standard errors to give the user a better understanding of model uncertainty and goodness of fit. The spatial distribution of model standard errors is of particular interest because it provides an assessment of the design of the well monitoring network. Areas of low monitoring density will have larger model standard errors. This not only informs the user that the predictions in this area need to be interpreted with care but also identifies potential locations where the construction of new monitoring wells would improve conceptual understanding of a site, and project decision-making. Model standard errors could also be used in the calculation of the solute plume characteristics mentioned above to provide a confidence interval on these quantities. 

Whilst simple to implement, the substitution of non-detect concentration values is not without its disadavanatages as discussed by \cite{HelselNADABook}. These are partly mitigated in GWSDAT by offering the `worse case' scenario of substitution with the full detection limit as opposed to the usual value of half the detection limit. However, the occurence of different detection limits for the same solute (perhaps because different laboratories were used during the course of a long-term monitoring programme) is still troublesome as substitution with any constant fraction leads to an apparent trend in concentrations. The authors are currently researching more sophisticated censored regression techniques to handle non-detect data in the spatiotemporal modelling framework.

\section*{Acknowledgements}

This work was funded by Shell Global Solutions (UK) Ltd. The authors acknowledge contributions from numerous colleagues to the development of GWSDAT: Dr Matthew Lahvis, Dr George Devaull, Dan Walsh, Curtis Stanley, 
and Professor Jonathan Smith of Shell Projects \& Technology HSE Technology; Philip Jonathan of Projects \& Technology - Analytical Services: Statistics \& Chemometrics; Ewan Crawford, of Glasgow University, Scotland, UK.
The views expressed are those of the authors and may not reflect the policy or position of Royal Dutch Shell plc.


\appendix

\section{Description of statistical modelling techniques} \label{ModellingDescription}

\subsection{Well trend plot smoother}\label{smregressCalc}

The well trend plot smoother is fitted using a nonparametric method called local linear regression. This involves solving locally the least squares problem:

\begin{equation}\label{locallinreg}
\textrm{min}_{\alpha , \beta} \sum_{i}^{n}\left\{ y_i - \alpha - \beta (x_i-x) \right\}^2 w(x_i-x; h)
\end{equation} 

\noindent where $w(x_i-x; h)$ is a weight function with parameter $h$. The weight function gives the most weight to the data points nearest the point of estimation 
and the least weight to the data points that are furthest away. For the weight function GWSDAT uses a normally-distributed probability density function with standard deviation $h$. 
Local linear regression is deployed in GWSDAT using the R package \emph{sm} \citep{sm1,smbook} and $h$ is selected using the method published in \cite{Hurvichetal}. 

\subsection{Groundwater flow estimation}\label{GWCalc}
Vectors of groundwater flow strength and direction are estimated using the well coordinates and recorded groundwater elevations. The model is based on the simple premise that local 
groundwater flow will follow the local direction of steepest descent (hydraulic gradient). 

For a given well, a linear plane is fitted to the local groundwater level data:

\begin{equation}\label{GWLinSurface}
L_i=a+bx_i+cy_i+\epsilon_i
\end{equation} 
where $L_i$ represents the groundwater level at location $(x_i,y_i)$. Local data is defined as the neighbouring wells 
as given by a Delaunay triangulation \citep{DelRef,deldir} of the monitoring well locations.
The gradient of this linear surface in both x and y directions is given by the coefficients $b$ and $c$. Estimated direction of flow is given by: 

\begin{equation}
\theta = \tan^{-1}\left (\frac{c}{b} \right)
\end{equation} 
and the relative hydraulic gradient (a measure of relative flow velocity) is given by 

\begin{equation}
R=\sqrt{b^2+c^2}
\end{equation} 

For any given model output interval this algorithm is applied to each and every well where a groundwater elevation has been recorded.

\subsection{Spatiotemporal solute concentration smoother}\label{STMathSmoother}

The spatiotemporal solute concentration smoother is fitted using a non parametric regression technique known as Penalised Splines \citep{Eilers92,Eilers96}.  A full and detailed explanation of  applying this statistical method to groundwater monitoring data is the subject of another paper \citep{Eversetal2013}.
However, the following outlines some of the most important aspects for the purposes of GWSDAT.

\vspace{0.5cm}
\noindent
Let $y_i$ be the natural log solute concentration at   $\boldsymbol{x_i}  =(x_{i1}, x_{i2}, x_{i3})$
where $x_{i1}$ and $x_{i2}$ stand for the spatial coordinates of the well and $x_{i3}$
represents the corresponding time point for the \textit{i-th}  observation with
$i = 1, \ldots, n$.\, We start by modelling the solute concentration as

\begin{equation}
\label{E:int}
\displaystyle   y_i = \sum_{j=1}^m\, b_j(\boldsymbol{x_i})\alpha_j + \epsilon_i
\end{equation}

\noindent
where the $b_j$, $j = 1, \ldots, m$
are $m$  B-Spline basis functions, generally second or third order polynomials \citep{Eilers96}. The measurement errors $\epsilon_i$'s are assumed to be iid normally distributed with zero mean and variance  $\sigma^2$.
Rewriting equation (\ref{E:int}) in the more compact matrix notation leads to

\begin{equation}
\bf{y =B(x) \boldsymbol\alpha +\boldsymbol\epsilon}
\end{equation}

The traditional ordinary least squares approach is to minimize the objective function  $\mathrm{S}\left(\boldsymbol\alpha\right) = \Vert
\boldsymbol\epsilon \Vert^2 = \Vert \boldsymbol y - \boldsymbol B(\boldsymbol x) \boldsymbol\alpha\Vert^2$. The well known major disadvantage of this approach is its propensity to overfit data leading to under smoothness in model predictions. 
To overcome this hurdle, the objective function is modified with the addition of a term that penalises the finite differences of the coefficients of adjacent B-splines.
The objective function now takes the form $\mathrm{S}\left(\boldsymbol\alpha\right)
  = \Vert \boldsymbol y - \boldsymbol B(\boldsymbol x) \boldsymbol\alpha\Vert^2 + \lambda \Vert D_d \alpha \Vert^2$ 
where $ D_d$ is a matrix such that $ D_d =  \Delta ^d$, the d-th differences of $\alpha$, and $\lambda$ is a nonnegative
tuning parameter.

By minimising the new objective function for a given value of $\lambda$,\, we obtain
the estimator of the parameters
  $\boldsymbol{\hat{\alpha}} = \left(\boldsymbol B'\boldsymbol B + \lambda \boldsymbol D'_d\boldsymbol D_d\right)^{-1}\boldsymbol B'\boldsymbol y$.\,
Note that when  $\lambda=0$, we have the standard ordinary least squares estimate for  $\boldsymbol{\hat{\alpha}}$.

Optimal selection of the penalisation parameter $\lambda$ is a subtle and important matter. A value which is too small leads to `overfitting', i.e. capturing the noise in the data. Conversely, a value which is too large leads to over smoothing of the data, i.e. `underfitting'. 
Several criteria have been traditionally proposed  (e.g. \cite{Hurvichetal,WoodB}) but the authors tackled this problem using a Bayesian modelling framework which is detailed in \cite{Eversetal2013}.

\bibliographystyle{elsarticle-harv}

\end{document}